# Direct imaging of a Berry curvature nematic state in a spin-compensated magnet

Weihang Lu,[1] Camron Farhang,[1] Yuchuan Yao,[2] Pratap Pal,[2] Hao Zhang,[4] Shaofeng Han,[1] Shi-Zeng Lin,[3] Chang-Beom Eom,[2] and Jing Xia[1,*]

[1] Department of Physics and Astronomy, University of California, Irvine, Irvine, CA 92697, USA

[2] Department of Materials Science and Engineering, University of Wisconsin-Madison, WI 53706, USA.

[3] T-4 & Center for Integrated Nanotechnologies (CINT), Los Alamos National Laboratory, Los Alamos, NM 87545, USA

[4] Department of Physics and Astronomy, The University of Tennessee, Knoxville, Tennessee 37996, USA

* xia.jing@uci.edu

**Abstract**

Density waves conventionally describe the periodic modulation of charge or spin, yet the spatial modulation of electronic geometry has remained elusive. Here, we report subtle micrometer-scale spatial modulations of the magneto-optical Kerr signal in the noncollinear antiferromagnet $Mn_3NiN$ with compensated spins, consistent with a magnetic-field-induced Berry curvature density wave . These Berry curvature modulations exhibit orientations unpinned from the crystal lattice, forming a nematic state that spontaneously breaks rotational symmetry. We attribute this spatial instability to field-induced spatial variations of the spin texture driven by competing magnetic interactions. This discovery unveils a new class of collective order in spin-compensated magnets mediated by the geometric phase of the wavefunction itself. Its wavelength is controlled by chemical doping and its amplitude by magnetic field, providing concrete tuning knobs for antiferromagnetic and altermagnetic spintronics.

## Introduction

Density waves (such as charge density wave (CDW) [1], spin density wave (SDW) [2], pair-density wave [3], and exciton density wave [4]) play a central role in understanding quantum materials because they represent collective electronic ordering that emerges from strong interactions between electrons, lattice vibrations, and magnetic degrees of freedom. Just as a CDW modulates the scalar charge distribution $\rho(r)$ and a SDW modulates the vector spin density $S(r)$, a Berry-curvature density wave would fill a critical void in this hierarchy by representing the periodic spatial modulation of the electronic topology itself, specifically, the momentum-space Berry curvature $\Omega(r)$ [5,6] projected into real space. When density waves form, they break fundamental symmetries of the system, such as translational [1] and rotational symmetry [7,8], leading to new ground states with distinct electronic, optical, and magnetic properties. For example, in strongly correlated superconductors like cuprates and iron-based high temperature superconductors, spin and charge density waves, some induced by a magnetic field [9], are strongly intertwined [10,11] with their superconductivity, where Cooper pair amplitude could also vary periodically in space establishing a pair density wave [12]. As another example, in the fractional quantum Hall system under a high magnetic field at half-filled Landau levels, electrons self-organize into an anisotropic "stripe" phase [13] where the charge density modulates in one direction, breaking rotational symmetry and leading to strongly direction-dependent transport. The presence of these density waves often signals competing or intertwined phases [10], such as superconductivity, Mott insulation, or topological order, making them an essential probe of the underlying quantum correlations. By tuning external parameters like pressure, temperature, or doping, transitions can be driven between density-wave states and other exotic phases [14–16], offering a powerful platform for discovering and engineering emergent phenomena in quantum materials.

Berry curvature [5,6] is a geometric property of electronic wavefunctions that acts like a pseudomagnetic field in momentum space, governing electron dynamics. It gives rise to topological effects such as the anomalous Hall effect [17] and quantized Hall plateaus [18–20] via the Chern number. It plays a central role in spin-compensated magnets, such as noncollinear antiferromagnets (AFM) [21] and altermagnets (AM) [22,23]. Despite negligible magnetization, large anomalous Hall effects (AHE) are observed in AFM [24] and in AM [25,26] driven purely by Berry curvature, offering promising platforms for ultrafast AFM and AM spintronics with minimal stray fields, field robustness, and fast switching [27–30]. Magneto-optic Kerr effect (MOKE) [31] and X-ray magnetic circular dichroism (XMCD) [23] have been used to directly image the Berry curvature in AFM $Mn_3Sn$ [31] and AM MnTe [23] respectively. These advances raise the intriguing possibility of observing spatial modulations of the Berry curvature, a Berry-curvature density wave, in spin-compensated magnets. Equally important is understanding how such Berry curvature modulations respond to external perturbations such as magnetic fields, which is crucial for advancing AM and AFM spintronic technologies. It is well known that competing magnetic interactions can stabilize long-wavelength modulated spin textures, such as spirals or magnetic skyrmions. Because the real-space spin texture is closely linked to the momentum-space Berry curvature, noncollinear antiferromagnets provide a promising materials platform to search for Berry-curvature density waves. To resolve the spatial modulation of the Berry curvature, it is crucial that the optical beam size in MOKE imaging be smaller than the period of the spin texture, yet larger than the atomic lattice constant.

Here using precision MOKE imaging, we discover that magnetic fields induce micron-scale spatial modulations in the otherwise uniform Berry curvature of single crystal epitaxial films of AFM $Mn_3NiN$ [32–35], forming Berry curvature density waves. $Mn_3NiN$ belongs to the family of noncollinear antiferromagnets including $Mn_3Sn$ [24], $Mn_3Ge$ [36], and $Mn_3Pd$ [37](*37*) [37] that exhibit noncollinear spin textures [21]. In the $\Gamma_{4g}$ spin texture of $Mn_3NiN$ [33] (Fig.1a), the Mn moments (blue arrows) form a triangular arrangement within the (111) kagome plane (blue triangle). This spin texture generates a nonzero Berry curvature $\Omega(k)$ in momentum space [21], which is a geometric property of Bloch electrons and can be represented as a pseudovector $\overrightarrow{\Omega_n}$ (red arrow in Fig.1a) oriented perpendicular to the kagome plane. Note that there are eight symmetry-equivalent $\overrightarrow{\Omega_n}$ domains corresponding to eight equivalent kagome planes [32–35], with four of them having identical positive z-component $\Omega_z$ , and the other four sharing a negative z-component $-\Omega_z$. In the presence of a Berry curvature modulation, the uniform pseudovector $\overrightarrow{\Omega_n}$ in a single domain becomes modulated in space, forming a pattern $\overrightarrow{\Omega_n}(\vec{r})$ as shown in Fig.1b with wavelength $l$. Experimentally, Berry curvature manifests in both the anomalous Hall effect (AHE) and the magneto-optical Kerr effect (MOKE) in these materials. The intrinsic anomalous Hall conductivity is given by: $\sigma_{xy}= \frac{-e^2}{2\pi^3\hbar}\int_{BZ} dk\, \Omega_z(k)$ (eq.1), where $\sigma_{xy}$ is the anomalous Hall conductivity, $\Omega_z(k)$ is the z-component of the total momentum-resolved Berry curvature. The polar MOKE signal $\theta_K(\omega)$ at optical frequency $\omega$: $\theta_K(\omega) = \mathrm{Re}\left[\frac{-\sigma_{xy}(\omega)}{\sigma_{xx}(\omega)\sqrt{1+i\left(\frac{4\pi}{\omega}\right)\sigma_{xx}(\omega)}}\right]$ (eq.2) provides a spatially resolved probe of the Berry curvature's z-component $\Omega_z$ [38]. Note that its physics origin is different from the SOC-free MOKE from real-space Berry curvature that we have recently observed in non-coplanar antiferromagnet $Co_{1/3}TaS_2$ [39].

**Results**

The 40 $nm$ thick single crystal epitaxial (001) $Mn_3NiN$ films used in this study were grown on (001)-oriented $(La_{0.3}Sr_{0.7})(Al_{0.65}Ta_{0.35})O_3$ (LSAT) substrates via sputtering. Their detailed structural, transport, and magnetic characterizations have been reported elsewhere [32,40], with the antiferromagnetic transition Néel temperature $T_N\sim250\ K$. The magnetic (001) neutron diffraction intensity of a similar $Mn_3NiN$ film emerges sharply below $T_N$, and remains nearly constant down to 2 $K$, indicating a fully developed temperature-invariant AFM order parameter [32]. A very small net moment $m_{net}$~0.008 $\mu_B/Mn$, due to spin-orbit coupling (SOC), enables magnetic field ($B$) switching of the chirality of the $\Gamma_{4g}$ AFM spin texture through Zeeman coupling. This chirality switching is accompanied with sign reversals of the Berry curvature $\Omega_z$, net moment $m_{net}$, AHE signal $\sigma_{xy}$, and MOKE signal $\theta_K$, and it is important to note that the vanishing moment $m_{net}$ is far too small to produce the observed large AHE and MOKE: they are instead induced by the Berry curvature [21,31,38,40]. Polar MOKE measurements reported here are performed using a zero-loop Sagnac interferometer microscope [41–44] in the polar geometry (Fig.1e) operating at the 1550 $nm$ telecommunication wavelength. Its superior 10 $nrad$ MOKE precision stems from its unique design that exclusively detects microscopic time-reversal symmetry breaking (TRSB) while rejecting non-TRSB effects such as anisotropy and vibrations. In a recent MOKE study [40] at the same 1550 $nm$ wavelength, we have demonstrated a temperature-invariant spontaneous MOKE response in $Mn_3NiN$ solely from the intrinsic Berry curvature contribution, which is in sharp contrast to the strongly temperature-dependent AHE that is dominated by skew scattering. As $\sigma_{xx}(\omega)$ at 1550 $nm$ is largely temperature independent in $Mn_3NiN$ [40], $\theta_K(\omega)$ scales with $\sigma_{xy}(\omega)$ (eq.2), which in turn scales with the Berry curvature integral (eq.1), allowing the imaging of Berry curvatures through spatially-resolved MOKE measurement: $\Omega_n \propto \theta_K$.

In sharp contrast to the smooth Hall effect in bulk transport measurement we reported earlier [32], the formation of a density wave is clearly visible in the MOKE image of a $Mn_3NiN$ film taken at $T = 2\ K$ and $B = 9\ T$ (Fig.1c). The horizontal and vertical ripples in the Kerr signal $\theta_K$ exhibit amplitudes of 2 to 4 $\mu rad$, which is up to 2% of the full 150 $\mu rad$ Kerr signal. It is noted that the Kerr signal would instead change from 150 $\mu rad$ to -150 $\mu rad$ across a chirality domain wall, which is 100 times larger than the observed subtle modulation. These ripples disappear completely either when the magnetic field is removed (Fig.2b), or when it is heated above the Néel temperature (Supplementary Fig.6). These ripples correspond to magnetic-field-induced spatial modulations of $\Omega_n$: Berry curvature density waves. It is important to note that this density wave spontaneously breaks both in-plane translational and rotational symmetries, as the applied magnetic field is oriented out of plane. Interestingly, several regions in Fig.1c display coexisting density waves with nearly orthogonal orientations. The intersection angle between these wave patterns varies continuously from 90° to 150°, as evident in the two-dimensional Fourier transform (Fig.1d), where the red and blue rectangles mark vertically and horizontally oriented stripes, respectively. The spread of orientations within the blue region indicates that the stripe directions are not locked to the $Mn_3NiN$ crystal's threefold rotational symmetry separated by 120°, but rather emerge spontaneously, breaking the rotational symmetry of the lattice (nematicity), likely directed by subtle local symmetry-breaking fields.

The magnetic field is identified as the driving force behind the emergence of the observed Berry curvature density wave. Fig.2a-c show MOKE images of the Berry curvature at $T = 2\,K$ during a magnetic field sweep: $B = 9\,T, 0\,T$, and $-9\,T$. The coercive field of $Mn_3NiN$ films is known to be massive below $150\,K$ [32,40], so $\pm 9\,T$ is not enough to reverse the chirality of the $\Gamma_{4g}$ antiferromagnetic spin texture (Supplementary Fig. 1). However, the applied field does modify the average Berry curvature, shifting the mean Kerr signal $\theta_K$ from 138 $\mu rad$ at zero-field to 158 and 68 $\mu rad$ at $9\,T$ and $-9\,T$ respectively. Superimposed on this average Berry curvature, clear density wave patterns appear at $B = \pm 9\,T$ (Fig.2a, c) but vanish completely $B = 0$ (Fig.2b). Although the density wave patterns at $B = -9\,T$ and $+9\,T$ broadly mirror each other, significant intrinsic asymmetries are observed, such as the amplitude discrepancy in the upper regions. These deviations (2 to 4 $\mu rad$) are two orders of magnitude larger than the $10\,nrad$ precision of the Sagnac interferometer, demonstrating that the density wave pattern is not strictly bound by static pinning centers. Instead, the spatial texture shifts as the magnetic field crosses zero (Supplementary Fig. 7: line-cut during $B = 9\,T \rightarrow 0\,T \rightarrow -9T$). This behavior is characteristic of a soft, collective mode that spontaneously reconfigures upon field reversal. Conversely, the pattern remains static when the field direction is unchanged, as observed during a $3\,T \rightarrow 6\,T \rightarrow 9\,T$ sweep (Supplementary Fig. 8).

At least two distinct sets of density waves are observed in both Fig.2a ($B = 9\,T$) and Fig.2c ($B = -9\,T$). In Fig.2a, the upper region is dominated by vertically oriented density waves, while the lower region exhibits horizontally oriented ones, with both patterns coexisting in the middle. To analyze these density waves in detail, the top and bottom regions of Fig.2a are extracted and shown in Fig.2d and Fig.2f, respectively. The corresponding x- and y- line profiles in Fig.2e and Fig.2g reveal nearly identical characteristics: both waves have an amplitude of approximately 2 $\mu rad$ and a period of 2.7 $\mu m$, despite their orthogonal orientations. All MOKE images obtained in this study consistently demonstrate that even though the geometric pattern changes over thermal cycling and magnetic field reversal, their amplitude and period are reproducible, indicating that these two key parameters are determined by the intrinsic properties of the antiferromagnet and by external parameters such as the applied magnetic field.

Additional MOKE imaging within the antiferromagnetic phase, performed in the same top region of Fig.2a under varying magnetic fields and temperatures, reveals that the amplitude of the Berry curvature modulation scales with the applied magnetic field $B_z$, while its period remains essentially constant. A phase diagram summarizing these results is shown in Fig.3a, where each solid data point represents the measured density-wave amplitude extracted from a MOKE image taken at the corresponding temperature (T) and field (B) (Supplementary Figs. 2-5). Representative MOKE images obtained at $T = 150\,K$ under magnetic fields of 9, 5, 2, and $0\,T$ are presented in Fig.3b, plotted as deviations $(\theta_K - \theta_{mean})$ from the average Kerr signal $\theta_{mean}$ for clarity. The spatial profile of the Berry curvature modulation changes significantly with temperature at a constant magnetic field (between $2\ K$ and $210\ K$, see Supplementary Figs. 2-5), further supporting the picture of a soft, spontaneously reconfigurable collective mode.

The density-wave amplitude increases systematically with magnetic field, yet its wavelength remains nearly unchanged. A similar trend is observed in the sequence recorded at $T = 2\ K$ (Fig.3c), confirming the field-scaling and field-independent periodicity of the Berry curvature modulation. The density-wave periodicity is found instead to correlate with chemical doping. As shown in Fig.4a, MOKE imaging of a slightly nitrogen-deficient $Mn_3NiN$ ($40\ nm$) film at $T = 235\ K$ and $B = 9\ T$ reveals that anti-doping more than doubles the density-wave period, clearly visible in the noise-filtered image in Fig.4b (raw data presented in Supplementary Fig. 9). The spontaneous MOKE signal vanishes completely when the sample is heated above the Néel temperature $T_N \sim 250\ K$ (Supplementary Fig. 6).

Unlike in two-dimensional electron systems where the nematic vector aligns with the host GaAs/AlGaAs $\langle 110 \rangle$ axis [13], or in the kagome superconductor $CsV_3Sb_5$ where it aligns with the kagome lattice [45], we observe that the direction of the Berry curvature modulation varies by up to 10° across thermal cycles. This variation demonstrates that the nematic vector is not entirely pinned by either the $Mn_3NiN$ lattice or the $(La_{0.3}Sr_{0.7})(Al_{0.65}Ta_{0.35})O_3$ (LSAT) substrate, suggesting a decoupling that is expected given the modulation's long wavelength ($\mu m$).

**Discussion**

Before interpreting the observed MOKE modulation as a Berry curvature density wave, we consider and rule out alternative explanations. First, domain contrast: polar MOKE contrast between $\Gamma_{4g}$ domains with positive/negative z-components of the Berry curvature ($\pm\Omega_z$) would be $\pm 150\ \mu rad$, which is much larger than the $\mu rad$-level subtle modulation, ruling out magnetic domain as the explanation. Second, local strain: strain fields are known to be present in epitaxial $Mn_3NiN$ [46]. However, strain-induced MOKE modulations would be present even at zero field and would not scale with applied field, in contrast to our observations.

The observation of coexisting, orthogonally oriented density waves (Fig. 2a, c), which are notably decoupled from the threefold (120°) rotational symmetry of the $Mn_3NiN$ lattice, serves as a hallmark of spontaneous rotational symmetry breaking. This identifies the phenomenon as a Berry curvature nematic state, where the electronic geometry self-organizes into a directional texture unpinned by the rigid crystal manifold. Analogous to the nematic stripe phase in fractional quantum Hall systems [13,7] and electronic nematicity in correlated superconductors [10,11], this state marks a regime where the collective modulation of the geometric phase breaks the point-group symmetry of the lattice.

The wavelength of the observed Berry curvature density wave is much larger than the underlying lattice constant, in sharp contrast to the much shorter wavelengths of conventional commensurate or incommensurate charge and spin density waves [1]. Unlike conventional density waves, which originate from electronic instabilities driven by the Fermi wave vector and typically manifest on the atomic scale, the Berry-curvature density wave observed here is a mesoscopic phenomenon. Its formation is driven by competing magnetic interactions that modulate the spin texture over micrometer length scales. Consequently, the resulting density wave differ fundamentally from standard CDWs or SDWs: they are collective modes of the geometric phase that emerge at wavelengths thousands of times larger than the unit cell. Consistent with its long wavelength, the orientation or location of the modulation is not pinned to the crystal symmetry: they shift with thermal cycles and magnetic field hysteresis. Both features resemble the nematic state of the two-dimensional electron gas in fractional quantum Hall systems [13,7], where the two-dimensional electronic gas is only weakly coupled to the host GaAs lattice. In the noncollinear antiferromagnet $Mn_3NiN$, although the orientations of the three Mn spins within each unit cell are fixed by the $Mn_3NiN$ lattice, collective ripples of many such spin textures are effectively free from lattice pinning, forming a long-wavelength, freely oriented Berry curvature density wave.

In the following, we present an effective minimalistic model, incorporating the coupling between conduction electrons and localized spins with competing magnetic interactions, to capture the main features of the observed Berry-curvature density wave. At zero magnetic field, the spins in $Mn_3NiN$ form a 120° coplanar noncollinear order (Fig.1a) within each triangular unit of the kagome plane [47,33]. When an out-of-plane magnetic field $B_z$ is applied, the spins gradually cant toward the field direction, becoming slightly non-coplanar. The conduction electrons mediate a Ruderman-Kittel-Kasuya-Yosida (RKKY) [48] interaction between localized moments, which is generally competing in nature. It has been shown that such competing magnetic interactions in the canted phase can lead to a spatial modulation of the canting angle [49,50]. An example of such an effective spin Hamiltonian producing this modulated canted phase is presented in the Supplementary Information. As illustrated in Fig.4c, the local canting angle is spatially modulated as a function of distance. A minimal Hamiltonian that includes both conduction electrons and localized spins and spin orbital coupling is given in the method section. The coupling between conduction electrons and the spin order in the presence of spin orbit coupling generates a finite Berry curvature even in the pristine uncanted phase. Experimentally, the observed Berry-curvature modulation has a period on the order of 10 $\mu m$, which is much longer than the atomic lattice constant. This justifies the usage of the adiabatic approximation in evaluating the Berry curvature, i.e., computing it from the Hamiltonian for each local canting angle. The results, shown in Fig.4d, e, reveal that the canting-angle modulation indeed gives rise to a spatially modulated Berry curvature consistent with experimental observations. The modulation period depends on the RKKY interaction, which is independent of the magnetic field and in turn is controlled by the conduction-electron density.

This theoretical model is consistent with the observed doubling of density wave period in the slightly nitrogen-deficient $Mn_3NiN$ (Fig.4b). Both the canting angle and the amplitude of its modulation increase with field [38]. Because the spatial modulation in the spin texture is determined by the competing RKKY interaction, which depends on the Fermi surface geometry, this result demonstrates that the Berry-curvature density wave can be deterministically tuned via the chemical potential. This tunability enables engineering control of Berry curvature for spintronic applications. The decoupling of the model from the underlying lattice is consistent with the coexisting, crossed density waves (Fig.2a, c) that emerge from nearly degenerate in-plane modulation directions. The linear scaling of amplitude with $Bz$ (Fig.3a) is consistent with this theoretical model, in which the canting angle (and therefore its spatial modulation amplitude) grows linearly with field in the weak-canting limit.

Based on this model, more complex Berry-curvature patterns can emerge. Superposing multiple helices can generate topological textures such as skyrmions [51] or merons [52,50] formed from three helices rotated by 120°. In Fig.4a, two nearly perpendicular Berry-curvature density waves interact to produce a checkerboard-like pattern. The above minimal model is expected to apply broadly to other noncollinear antiferromagnets such as $Mn_3Sn$ [24], $Mn_3Ge$ [36], and $Mn_3Pd$ [37], implying the possible existence of Berry-curvature density waves in these systems. Furthermore, while this study establishes the Berry-curvature density wave in a non-collinear antiferromagnet, the underlying mechanism may broadly apply to the emerging class of spin-compensated topological magnets, including altermagnets. In these systems, large Berry curvatures generated by crystal symmetries could similarly succumb to long-wavelength instabilities, even in the absence of net magnetization. Spintronics devices based on these materials have already been demonstrated, including $Mn_3Ge$-based spin-triplet supercurrent spin valves and SQUID [53], and $Mn_3Sn$-based volatile memory [54]. The micron-scale Berry-curvature density waves observed here are therefore highly relevant for antiferromagnetic and altermagnetic spintronics: controlled Berry-curvature ripples could be exploited as active functional elements, while in other cases they may need to be minimized, for instance through magnetic field shielding.

## Methods

**Film Growth:** Stoichiometric single crystal epitaxial 40 nm $Mn_3NiN$ thin films were grown on (001)-oriented $(La_{0.3}Sr_{0.7})(Al_{0.65}Ta_{0.35})O_3$ (LSAT) substrates at 650°C under an Ar/$N_2$ (78%:22%) mixed atmosphere of 10 mTorr by DC reactive magnetron sputtering using a stoichiometric $Mn_3Ni$ target, which was monitored by in-situ reflection high energy electron diffraction (RHEED).

**Sagnac polar MOKE measurements** are performed using a zero-loop fiber-optic Sagnac interferometer [55] microscope [43,44]. Upon reflection from the sample the nonreciprocal phase shift $\Delta\varphi$ between the two counterpropagating circularly polarized beams is twice the Kerr rotation $\Delta\varphi = 2\theta_K$. Both the optical beam and the applied magnetic fields are along the [001] crystallographic direction (z-direction, perpendicular to the film). The detailed operation of a Sagnac interferometer is described in the Supplementary Information.

**Modulated canting angle and Berry curvature calculations:** The modulated canting originates from competing magnetic interactions. The spin texture **S(r)** in $Mn_3NiN$ lies mainly in the $xy$ plane and is canted by an external field along $z$. The local spin energy $\mathcal{H}_s = \int d\mathbf{r}(A\, S_z^2 - B_z\, S_z)$ favors uniform canting, while the competing RKKY exchange $\mathcal{H}_{RKKY} = \int d\mathbf{r} d\mathbf{r}' J(\mathbf{r}-\mathbf{r}') S(\mathbf{r}) S(\mathbf{r}') = \int d\mathbf{q} J(\mathbf{q}) S(\mathbf{q}) S(-\mathbf{q})$, with $J(\mathbf{q})$ being minimal at a finite wave vector $\mathbf{Q}$, drives spatial modulation of spins. Adding an inter-sublattice antiferromagnetic term $\mathcal{H}_{AFM} = J_{AFM} \int d\mathbf{r} \big(\mathbf{S}_1(\mathbf{r}) \cdot \mathbf{S}_2(\mathbf{r}) + \mathbf{S}_2(\mathbf{r}) \cdot \mathbf{S}_3(\mathbf{r}) + \mathbf{S}_3(\mathbf{r}) \cdot \mathbf{S}_1(\mathbf{r})\big)$ stabilizes the 120° noncollinear order observed in $Mn_3NiN$. This modulated canting directly induces a real-space modulation of the local Berry curvature. Periodic variations in the canting angle alter the local electronic band structure, producing a Berry-curvature density wave with the same period as **Q**. The modulation wavelength is determined by the RKKY interaction, which depends on the conduction-electron density and Fermi-surface geometry. Details of the calculation is described in the Supplementary Information.

## Data availability

Source data are provided with this paper. They have been deposited in a figshare repository with (link TBA).

**Acknowledgements**

This project was supported by NSF award DMR-2419425 and the Gordon and Betty Moore Foundation EPiQS Initiative, Grant # GBMF10276 awarded to J.X.. C.B.E. acknowledges support for this research through a Vannevar Bush Faculty Fellowship (ONR N00014-20-1-2844) and the Gordon and Betty Moore Foundation's EPiQS Initiative, Grant GBMF9065 awarded to C.B.E.. S.-Z. L. is supported by the U.S. Department of Energy (DOE) National Nuclear Security Administration (NNSA) under Contract No. 89233218CNA000001 through the Laboratory Directed Research and Development (LDRD) Program and was performed, in part, at the Center for Integrated Nanotechnologies, an Office of Science User Facility operated for the DOE Office of Science, under user Proposals No. 2018BU0010 and No. 2018BU0083.

**Author Contributions**

J.X. conceived and supervised the project. W.L., C.F., S.H., and J.X. carried out the experimental measurements. Y.Y., P.P., and C.B.E. fabricated and characterized the samples. H.Z. and S.-Z. L. performed theoretical calculation. J.X. analyzed the data and drafted the paper with the input from all authors. All authors contributed to the discussion of the manuscript.

**Competing interests**

The authors declare no competing interest.

**Correspondence** and requests for materials should be addressed to Jing Xia, Email: xia.jing@uci.edu.

**Figures and Captions**

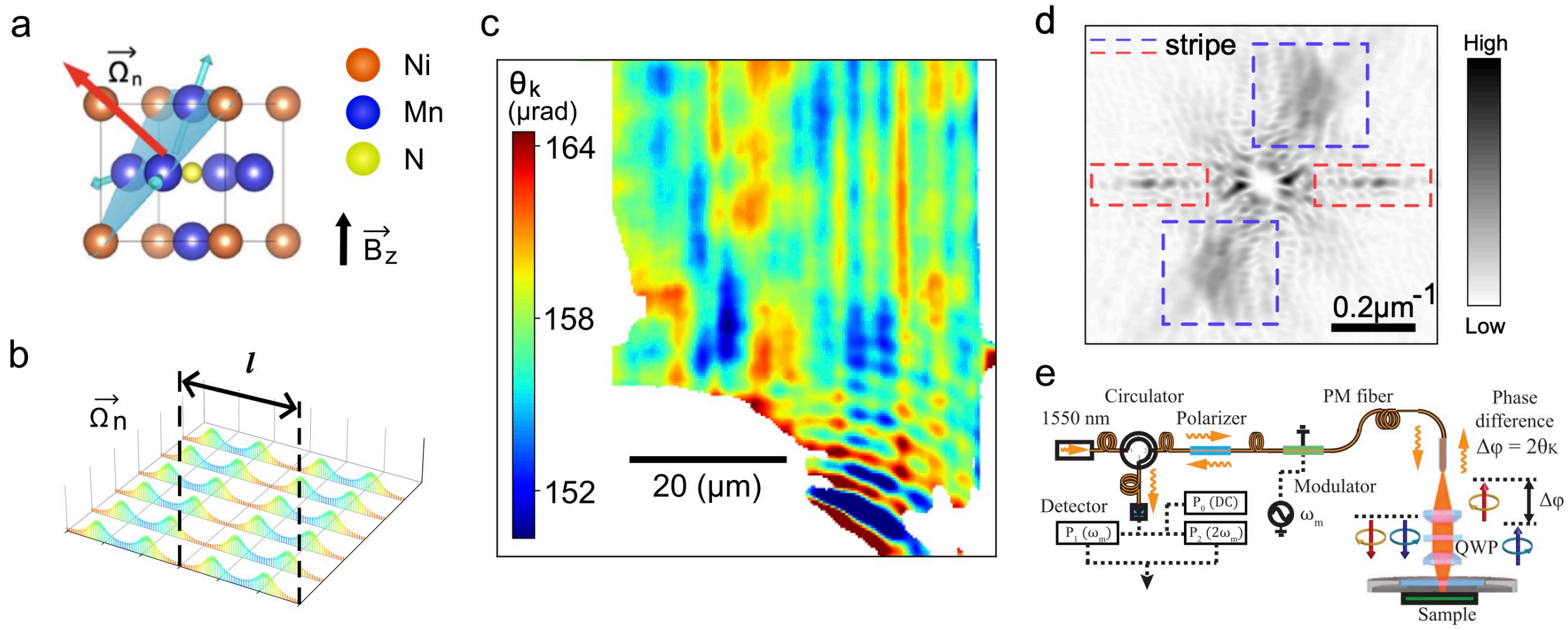

**Figure 1. Berry-curvature density wave in noncollinear antiferromagnet $Mn_3NiN$. a,** the noncollinear antiferromagnetic ($\Gamma_{4g}$) spin texture in $Mn_3NiN$ resulting in a non-zero Berry curvature pseudovector $\overrightarrow{\Omega_n}$ (red arrow) perpendicular to the kagome plane (blue triangle). Note there are eight symmetry-equivalent $\overrightarrow{\Omega_n}$ corresponding to eight equivalent kagome planes. Blue vectors represent Mn spins. **b,** Schematic of a real-space Berry curvature density wave. A "ripple" of the pseudovector $\overrightarrow{\Omega_n(r)}$ emerges in real space with a period $l$. **c,** MOKE image of the Berry curvature at T = 2 K and B = 9 T, showing two orthogonally oriented density waves. **d,** Corresponding 2D-FFT with red and blue squares marking the horizontally and vertically oriented density waves respectively. **e,** schematic of the Sagnac interferometer microscope for MOKE imaging.

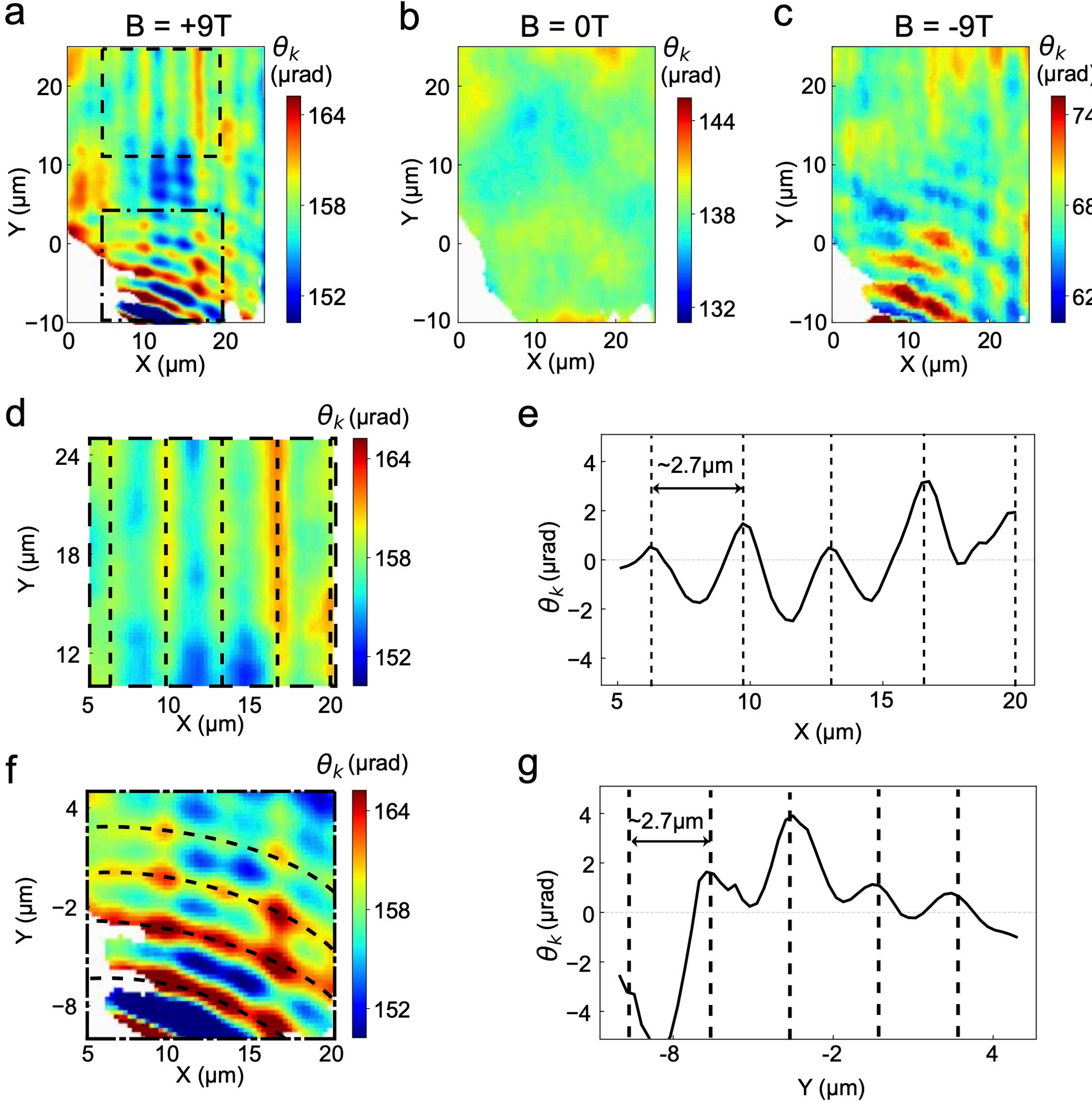

**Figure 2. Magnetic field drives Berry-curvature density wave. a, b, c,** MOKE images of Berry curvature at $2\ K$, taken with an applied magnetic field of $9\ T$, $0\ T$, and $-9\ T$ respectively. Note the emergence of density wave at $\pm 9\ T$, and its absence at zero field. **d, e,** MOKE image and x-line plot respectively of a region with x-direction Berry-curvature density wave taken at 2 K and 9 T. **f, g,** MOKE image and y-line plot respectively of a region with y-direction Berry-curvature density wave taken at 2 K and 9 T. Note the identical $2.7\mu m$ period for these two orthogonally oriented density waves.

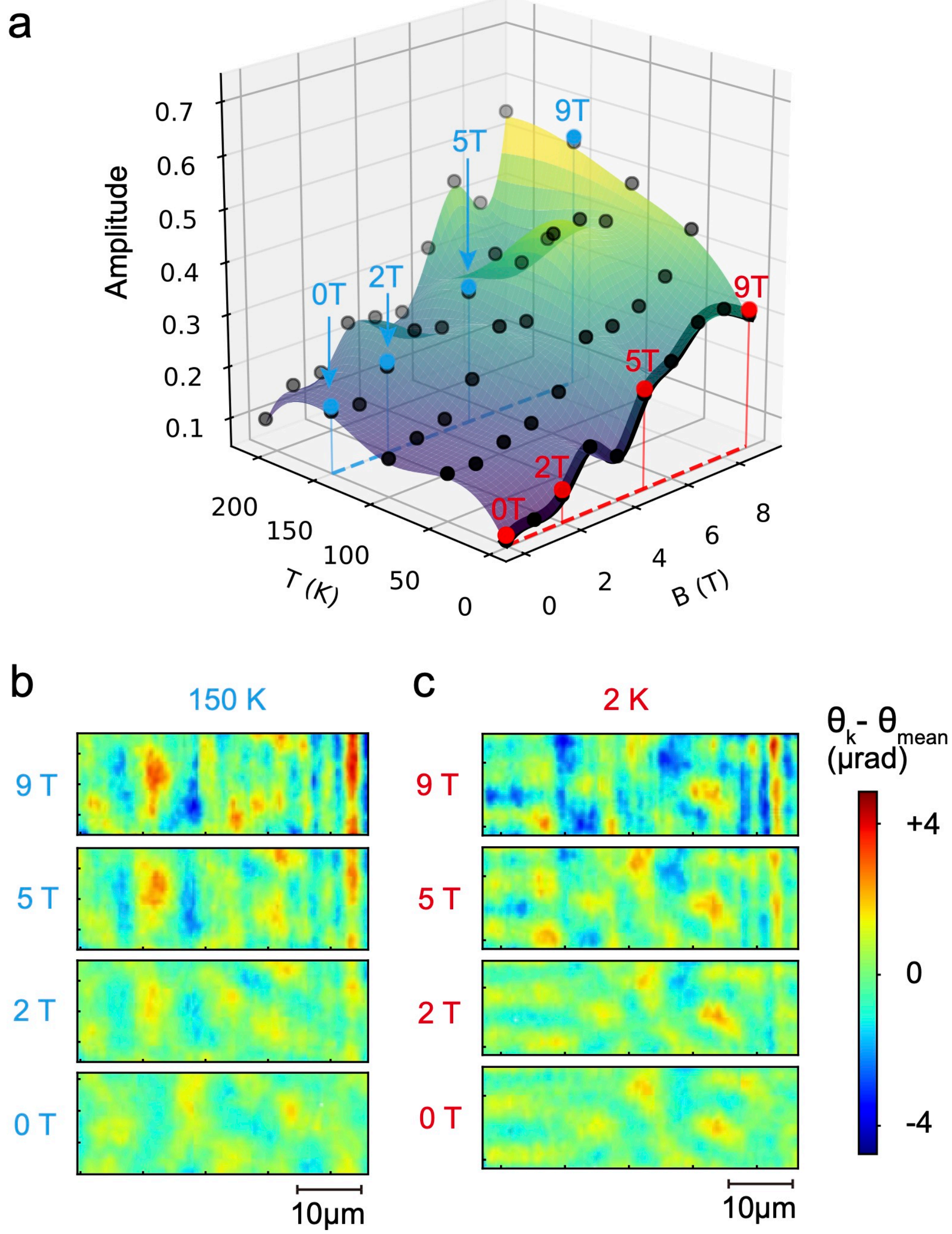

**Figure 3. Phase diagram of Berry-curvature density wave. a,** Amplitude of the Berry-curvature density wave as a function of temperature T and magnetic field B. **b, c,** MOKE images of Berry-curvature at representative magnetic fields of 9, 5, 2, and 0 $T$, taken at 150 $K$ and 2 $K$ respectively. They are plotted as the deviation ($\theta_K - \theta_{mean}$) from the average MOKE signal $\theta_{mean}$ for clarity.

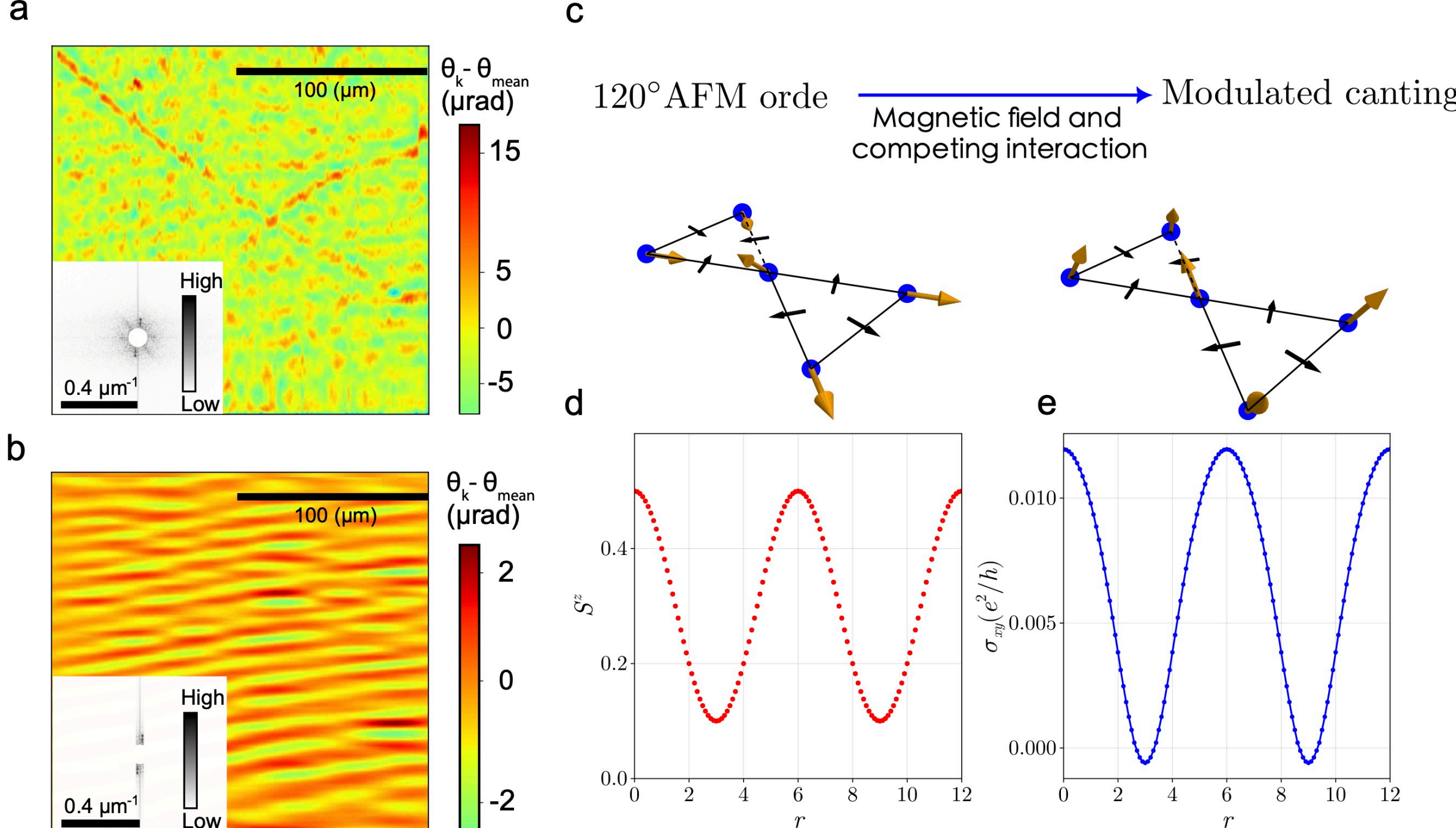

**Figure 4. Mechanism of Berry-curvature density wave. a,** Berry-curvature density wave imaged by MOKE in another $Mn_3NiN$ sample. Insets are Fourier transforms of the image. **b,** noise-filtered MOKE image (raw data presented in Supplementary Fig. 9), keeping only the horizontally oriented density wave for clarity. **c,** Schematics of the zero-field in-plane 120° magnetic order (left) and the modulated canted antiferromagnetic spin texture (right) induced by an external magnetic field and competing interactions in the Kagome plane. The black arrows indicate the symmetry-allowed spin–orbit vectors $n_{\alpha\beta}$ defined in the Supplementary Information. **d, e,** We use the ansatz, $S^z(\mathrm{r}) = \alpha_0 + \alpha \cos\left(\frac{2\pi r}{l}\right)$ to describe the modulated canting of spins. The theoretically calculated Hall conductivity $\sigma_{xy}$ as a function of distance $r$, showing the formation of a long wavelength density wave. See Supplementary Information for details of the calculations.